\documentclass[preprint,12pt]{elsarticle}

\usepackage{amssymb}
\usepackage{mathrsfs}
\usepackage[breaklinks,colorlinks,citecolor=green,urlcolor=blue,linkcolor=red]{hyperref}

\journal{Physics Letters B}

\begin{document}

\begin{frontmatter}



\title{A novel observable for $C\!P$ violation in multi-body decays and its application potential to charm and beauty meson decays}


\author{Zhen-Hua Zhang}
\ead{zhangzh@usc.edu.cn}
\address{School of Nuclear Science and Technology, University of South China,
            Hengyang, 421001,
            Hunan,
            China}


\begin{abstract}
A novel observable measuring the $C\!P$ asymmetry in multi-body decays of heavy mesons, which is called the forward-backward asymmetry induced $C\!P$ asymmetry (FBI-$C\!P$A), $A_{CP}^{FB}$, is introduced.
This  observable has the dual advantages that 1) it can isolate the $C\!P$ asymmetry associated with the interference of the $S$- and $P$-wave amplitude from that associated with the $S$- or $P$-wave amplitude alone; 2) it can effectively almost double the statistics comparing to the conventionally defined regional $C\!P$ asymmetry.
We also suggest to perform the measurements of FBI-$C\!P$A in some three-body decay channels of charm and beauty mesons.
\end{abstract}

\begin{keyword}
$C\!P$ violation \sep multi-body decay \sep heavy meson \sep B meson \sep D meson


\end{keyword}

\end{frontmatter}


\section{\label{sec:introduction}Introduction}

$C\!P$ violation is an important ingredient of the Standard Model of particle physics \cite{Kobayashi:1973fv}, also one of the necessary conditions for the dynamical generation of the baryonic asymmetry of the Universe \cite{Sakharov:1967dj}.
It was first discovered in the neutral kaon system in the year 1964 \cite{Christenson:1964fg}, and its discoveries in $B$ meson decay processes by the $B$ factories confirmed the Cabibbo-Kobayashi-Maskawa mechanism of SM \cite{Aubert:2001nu,Abe:2001xe,Abe:2004us,Aubert:2007mj,Aaij:2012kz,Aaij:2013iua,Aaij:2017ngy}.
Recently, $C\!P$ violation was also discovered in the charmed meson decay processes \cite{Aaij:2019kcg}.

Intensive studies on $C\!P$ violations in multi-body decays of beauty and charmed hadrons have been performed both theoretically \cite{Zhang:2013oqa,Bhattacharya:2013cvn,Bediaga:2013ela,Wang:2015ula,Cheng:2016shb,Klein:2017xti,Dedonder:2010fg,Cheng:2020ipp,Zhang:2020rol} and experimentally \cite{Aubert:2009av,Aaij:2013sfa,Aaij:2013bla,Aaij:2014iva,Aaij:2018bjo,Aaij:2019hzr,Aaij:2019jaq,Aaij:2020wil} during the last ten years.
One advantage for multi-body decays is that $C\!P$ violation can be studied through the phase space distribution of the decay, namely, the regional $C\!P$ asymmetries distributed in the phase space.
The total decay amplitudes can be expressed as a superposition of various amplitudes, which can allow the presence of different strong phases.
Because of the interference effects of these amplitudes, the regional $C\!P$ asymmetries in certain places of the phase space can be very large.
Up to now, the regional $C\!P$ asymmetry is one of the most important and extensively studied observables associated with $C\!P$ violation in multi-body decays, other than the integrated $C\!P$ asymmetry.
Although for four-body decay channels and baryon three-body decay channels, one can also study the $C\!P$ violation associated with triple product asymmetry \cite{Gronau:2011cf,Gronau:2015gha,Aaij:2016cla,Aaij:2019mmy}.

The disadvantage of the regional $C\!P$ asymmetry in multi-body decays is also obvious.
Once focusing on a small region of the phase space, the experimental study of regional $C\!P$ asymmetries will suffer from low statistics.

In this paper, other than the regional $C\!P$ asymmetry, we are going to introduce an observable to measure the $C\!P$ violation in multi-body decays of heavy mesons, which according to our analysis below, can almost effectively double the statistics comparing to the conventionally defined regional $C\!P$ asymmetries.
Furthermore, this observable could potentially promote the discovery of $C\!P$ violation in multi-body decays of beauty and charmed mesons.

\section{\label{sec:observable}\boldmath The Forward-Backward asymmetry induced $C\!P$ asymmetry}

Consider a multi-body decay $H\to h_1 h_2 h_3\cdots h_n$, where $H$ is a heavy meson, and $h_1$, $h_2$, $\cdots$, $h_n$ are light ones.
We will focus on the phase space in the vicinity of a $P$-wave intermediate resonance $X$, where, the decay will be dominated by the cascade decay $H\to X h_3\cdots h_n$, $X\to h_1 h_2$.
The part of the phase space which we focus on satisfies $(m_X-\sigma_X)^2<s_{12}<(m_X+\sigma_X)^2$, where $s_{12}$ is the invariant mass squared of $h_1$ and $h_2$, $m_X$ is the mass of $X$, $\sigma_X$ is of the same order with the decay width of $X$, $\Gamma_X$.
Let us denote the relative angle between the momenta of $h_1$ and $H$ in the rest frame of $h_1$ and $h_2$ system (hence, of $X$) as $\theta^\ast_{1}$. 
Then, the part of phase space  that we focus on can be further divided into two parts according to whether $\theta^\ast_{1}$ is larger or smaller than $\pi/2$.
An observable, describing the forward-backward asymmetry in these two parts of the phase space, can be defined as
\begin{equation}\label{Eq:defAFB}
A^{FB}_{H\to h_1 h_2 h_3\cdots h_n}= \frac{\Gamma_{H}(c_{\theta^\ast_{1}}>0)-\Gamma_{H}(c_{\theta^\ast_{1}}<0)}{\Gamma_{H}(c_{\theta^\ast_{1}}>0)+\Gamma_{H}(c_{\theta^\ast_{1}}<0)}.
\end{equation}
where $c_{\theta^\ast_{1}}\equiv\cos\theta^\ast_{1}$, $\Gamma_{H}(c_{\theta^\ast_{1}}> 0)$ and $\Gamma_{H}(c_{\theta^\ast_{1}}< 0)$ are the regional decay widths of $H\to h_1 h_2 h_3\cdots h_n$ in the aforementioned two part of the phase space.\footnote{
Note that $\Gamma_{H\to h_1 h_2 h_3 \cdots h_n}(c_{\theta^\ast_{1}}\gtrless  0)$ is in fact abbreviation for $\Gamma_{H\to h_1 h_2 h_3\cdots h_n}((m_X-\sigma_X)^2<s_{12}<(m_X+\sigma_X)^2, c_{\theta^\ast_{1}}\gtrless  0)$, where the constraint $(m_X-\sigma_X)^2<s_{12}<(m_X+\sigma_X)^2$ has been omitted.}

The nonzero of $A^{FB}_{H\to h_1 h_2 h_3\cdots h_n}$ indicates that the decay amplitude of $H\to h_1 h_2 h_3\cdots h_n$ in the region of phase space $(m_X-\sigma_X)^2<s_{12}<(m_X+\sigma_X)^2$ is not only just dominated by  the cascade decay $H\to X (\to h_1 h_2) h_3$, but other contributions, usually $S$-wave amplitude, are also comparable.
This can be seen as follows.
Suppose that the amplitude of $H\to h_1 h_2 h_3\cdots h_n$ in the region of phase space $(m_X-\sigma_X)^2<s_{12}<(m_X+\sigma_X)^2$ is dominated by the cascade decay $H\to X (\to h_1 h_2) h_3\cdots h_n$, plus an $S$-wave amplitude, so that it can be expressed as a coherent sum:
\begin{equation}
  \mathcal{M}_{H\to h_1 h_2 h_3\cdots h_n}=\mathcal{M}_{H\to X(\to h_1 h_2)h_3\cdots h_n}+\mathcal{M}_{S-{\rm wave}},
\end{equation}
where, the amplitude of the cascade decay $H\to X (\to h_1 h_2) h_3\cdots h_n$ can be parameterized as
$\mathcal{M}_{H\to X(\to h_1 h_2)h_3\cdots h_n}=a_P c_{\theta_{1}^\ast}$,\footnote{
Note that there is a Breit-Wigner factor ${1}/{(s_{12}-m_X^2+im_X\Gamma_X)}$ in $a_P$.}
 while the $S$-wave amplitude can be parameterized as
$\mathcal{M}_{S-{\rm wave}}=a_S$.
Then, the differential decay width will take the form
\begin{eqnarray}
  d\Gamma&\propto&\left|\mathcal{M}_{H\to h_1 h_2 h_3\cdots h_n}\right|^2 ds_{12}dc_{\theta_{1}^\ast}d\tau\nonumber\\
  &=&\left[|a_P|^2c_{\theta_{1}^\ast}^2+|a_S|^2+2\Re(a_Pa_S^\ast)c_{\theta_{1}^\ast}\right]ds_{12}dc_{\theta_{1}^\ast}d\tau,
\end{eqnarray}
where a Jacobi factor corresponding to the variable transformation from $s_{13}$ to $c_{\theta_{1}^\ast}$ is omitted.
After integrating over $ds_{12}$, $d\tau$ and $c_{\theta_{1}^\ast}$, one has
\begin{equation}\label{Eq:Gammacgl0}
  \Gamma_{H\to h_1 h_2 h_3\cdots h_n}(c_{\theta_{1}^\ast}\gtrless 0)\propto \left\langle \left|a_P\right|^2\right\rangle/3+\left\langle\left|\langle a_S\right|^2\right\rangle\pm\Re(\langle a_Pa_S^\ast\rangle),
\end{equation}
where the angle brackets represent the phase space integration over all variables --$s_{12}$ is integrated from $(m_X-\sigma_X)^2$ to $(m_X+\sigma_X)^2$-- except $c_{\theta_{1}^\ast}$, i.e., $\langle\cdots\rangle=\int_{(m_X-\sigma_X)^2}^{(m_X+\sigma_X)^2} (\cdots)ds_{12}d\tau$, $c_{\theta_{1}^\ast}$ is integrated from -1 to 0 or from 0 to 1, respectively.
By substituting Eq. (\ref{Eq:Gammacgl0}) into Eq. (\ref{Eq:defAFB}), the forward-backward asymmetry can be expressed as
\begin{equation}
  A^{FB}_{H\to h_1 h_2 h_3\cdots h_n}= \frac{\Re(\langle a_Pa_S^\ast\rangle)}{\left\langle\left| a_P\right|^2\right\rangle/3+\left\langle\left| a_S\right|^2\right\rangle}.
\end{equation}
From the above equation one can clearly see that the presence of both the $P$- and $S$-wave amplitudes $\mathcal{M}_{H\to X(\to h_1 h_2)h_3\cdots h_n}$ and $\mathcal{M}_{S-{\rm wave}}$ results in nonzero forward-backward asymmetry $A^{FB}_{H\to h_1 h_2 h_3\cdots h_n}$.
On the other hand, if only the $P$-wave amplitude $\mathcal{M}_{H\to X(\to h_1 h_2)h_3\cdots h_n}$ contributes, $A^{FB}_{H\to h_1 h_2 h_3\cdots h_n}$ would simply be zero.

Up to this point, nothing is mentioned about the $C\!P$ conjugate process $\bar{H}\to \bar{h}_1 \bar{h}_2 \bar{h}_3\cdots \bar{h}_n$, and hence, the $C\!P$ asymmetry.
It is easy to see that if $C\!P$ symmetry is respected, one would simply have $A^{FB}_{H\to h_1 h_2 h_3\cdots h_n}=A^{FB}_{\bar{H}\to \bar{h}_1 \bar{h}_2 \bar{h}_3\cdots \bar{h}_n}$.
On the other hand, if the $C\!P$ symmetry is violated, $A^{FB}_{H\to h_1 h_2 h_3\cdots h_n}$ and $A^{FB}_{\bar{H}\to \bar{h}_1 \bar{h}_2 \bar{h}_3\cdots \bar{h}_n}$ would not equal to each other.
Consequently, one can introduce a new observable measuring the $C\!P$ asymmetry of multi-body decay $H\to h_1 h_2 h_3\cdots h_n$, which will be called the forward-backward asymmetry induced $C\!P$ asymmetry (FBI-$C\!P$A) hereafter and is defined as
\begin{equation}
  A^{FB}_{C\!P}=\frac{1}{2}(A^{FB}_{H\to h_1 h_2 h_3\cdots h_n}-A^{FB}_{\bar{H}\to \bar{h}_1 \bar{h}_2 \bar{h}_3\cdots \bar{h}_n}).
\end{equation}
From the definition of FBI-$C\!P$A one can easily see that its nonzero value indeed represents the violation of $C\!P$.

\section{\label{sec:discussion}\boldmath Discussions on FBI-$C\!P$A}

One of the motivations for the introduction of $A^{FB}_{C\!P}$ can be explained as follows.
When the $S$-wave amplitude are comparable with the $P$-wave one in the vicinity of the resonance $X$, the regional $C\!P$ asymmetries for $c_{\theta^\ast_{1}}>0$ and $c_{\theta^\ast_{1}}<0$, which are conventionally defined as
\begin{equation}\label{regionACP1part}
A^{{\rm reg}}_{C\!P}(c_{\theta^{\ast}_{1}}\gtrless 0)=\frac{\Gamma_{H}(c_{\theta^\ast_{1}}\gtrless0)-\Gamma_{{\bar H}}(c_{\theta^\ast_{1}}\gtrless0)}{\Gamma_{H}(c_{\theta^\ast_{1}}\gtrless0)+\Gamma_{{\bar H}}(c_{\theta^\ast_{1}}\gtrless0)},
\end{equation}
are correlated with each other.
To see this, one just needs to reexpressed $A_{C\!P}(c_{\theta^\ast_{1}}\gtrless 0)$ as
\begin{equation}\label{regionACP1part3origin}
A^{{\rm reg}}_{C\!P}(c_{\theta^\ast_{1}}\gtrless 0)= \frac{\frac{1}{3}A_{C\!P}^P+\frac{\left\langle\left| {a}_S\right|^2\right\rangle+\left\langle\left| {\bar a}_S\right|^2\right\rangle}{\left\langle\left| {a}_P\right|^2\right\rangle+\left\langle\left| {\bar a}_P\right|^2\right\rangle}A_{C\!P}^S\pm\frac{\Re(\langle a_Pa_S^\ast\rangle-\langle{\bar a}_P{\bar a}_S^\ast\rangle)}{\left\langle\left| {a}_P\right|^2\right\rangle+\left\langle\left| {\bar a}_P\right|^2\right\rangle}}{\frac{1}{3}+\frac{\left\langle\left| {a}_S\right|^2\right\rangle+\left\langle\left| {\bar a}_S\right|^2\right\rangle}{\left\langle\left| {a}_P\right|^2\right\rangle+\left\langle\left| {\bar a}_P\right|^2\right\rangle}\pm\frac{\Re(\langle a_Pa_S^\ast\rangle+\langle{\bar a}_P{\bar a}_S^\ast\rangle)}{\left\langle\left| {a}_P\right|^2\right\rangle+\left\langle\left| {\bar a}_P\right|^2\right\rangle}},
\end{equation}
by substituting Eq. (\ref{Eq:Gammacgl0}) into Eq. (\ref{regionACP1part}),
where $A_{C\!P}^{S/P}=\frac{\left\langle\left| {a}_{S/P}\right|^2\right\rangle-\left\langle\left| {\bar a}_{S/P}\right|^2\right\rangle}{\left\langle\left| {a}_{S/P}\right|^2\right\rangle+\left\langle\left| {\bar a}_{S/P}\right|^2\right\rangle}$.
It can be seen that there are three terms in the numerator of the above equation, corresponding to three origins of the regional $C\!P$ asymmetry, $A_{C\!P}(c_{\theta^\ast_{1}}\gtrless 0)$: the $C\!P$ asymmetry associated with the $S$- and $P$-wave alone, and that associated with the interference effect between the $S$- and $P$-waves.
Among these three terms, the first two are the same for
$A_{C\!P}(c_{\theta^\ast_{1}}> 0)$ and $A_{C\!P}(c_{\theta^\ast_{1}}< 0)$, except for the difference in the denominator, while the last one changes signs.
It is easy to see that the last origin of $C\!P$ asymmetry for $A_{C\!P}(c_{\theta^\ast_{1}}\gtrless 0)$ is proportional to the sine of the relative strong angle between the $S$- and $P$-wave amplitudes, which, according to Watson's theorem \cite{Watson:1952ji}, comes from the final state interaction, so that it can be large because of its nonperturbative attribute.
As a consequence, the last term in the numerator of Eq. (\ref{regionACP1part3origin}) can be comparable with --some times it can even dominate over-- the first two terms, resulting in a substantial difference between $A^{{\rm reg}}_{C\!P}(c_{\theta^\ast_{1}}> 0)$ and $A^{{\rm reg}}_{C\!P}(c_{\theta^\ast_{1}}< 0)$.
In fact, it has a good chance that the signs of $A^{{\rm reg}}_{C\!P}(c_{\theta^\ast_{1}}> 0)$ and $A^{{\rm reg}}_{C\!P}(c_{\theta^\ast_{1}}< 0)$ are opposite because of the presence of the last term in the numerator of Eq. (\ref{regionACP1part3origin}).
Indeed, this kind of behaviour has already been observed in $B^\pm\to \pi^+\pi^- K^\pm$ and $B^\pm\to \pi^+\pi^- \pi^\pm$ \cite{Aaij:2014iva}, and has been studied extensively in the literature.
One interesting property of the newly defined FBI-$C\!P$A is that it is capable of isolating the $C\!P$ asymmetry associated with the interference of the $S$- and $P$-waves,
which can be seen by expressing $A_{C\!P}^{FB}$ as
\begin{equation}
  A^{FB}_{C\!P}=\frac{\Re(\langle a_Pa_S^\ast\rangle)}{\left\langle\left| a_P\right|^2\right\rangle/3+\left\langle\left| a_S\right|^2\right\rangle}-\frac{\Re(\langle {\bar a}_P{\bar a}_S^\ast\rangle)}{\left\langle\left| {\bar a}_P\right|^2\right\rangle/3+\left\langle\left| {\bar a}_S\right|^2\right\rangle}.
\end{equation}
It is this property which motivates the introduction of FBI-$C\!P$A. \footnote{One can see that in contrast to the conventionally defined $C\!P$ asymmetry, there are extra factors $1/\left({\left\langle\left| {a}_P\right|^2\right\rangle/3+\left\langle\left| { a}_S\right|^2\right\rangle}\right)$ and $1/\left({\left\langle\left| {\bar a}_P\right|^2\right\rangle/3+\left\langle\left| {\bar a}_S\right|^2\right\rangle}\right)$ in the above expression. Of course, when $CP$ violation is small, so that the above two factors are nearly equal, $A^{FB}_{C\!P}$ would be proportional to $\Re(\langle a_Pa_S^\ast\rangle)-\Re(\langle {\bar a}_P{\bar a}_S^\ast\rangle)$, just like the third term in the numerator of Eq. (\ref{regionACP1part3origin}) does.}

One important issue, which has to do with the integration interval of $s_{12}$, should be pointed out here.
In the above discussion, the integral of $s_{12}$ is performed symmetrically around the $X$ resonance, i.e. $(m_X-\sigma_X)^2<s_{12}<(m_X+\sigma_X)^2$.
This may result in some cancellation when obtaining the FBI-$C\!P$A or regional $C\!P$As.
This has to do with the fact that the interference term may flip its sign at $s_{12}\sim m_{X}^2$.
To see this in more detail, let us first write out the Breit-Wigner in the $P$-wave amplitude explicitly:
\begin{equation}
  a_P=\frac{\tilde{a}_P}{s_{12}-m_{X}^2+im_X\Gamma_X},
\end{equation}
where $\tilde{a}_P$ is introduced to isolate the Breit-Wigner factor from $a_P$,
so that the interference term in $A^{FB}$, $A_{CP}^{FB}$, and $A_{CP}^{\rm reg}$'s can be expressed as
\begin{equation}
\Re( a_Pa_S^\ast)=\frac{2c_{\theta_{13}^\ast}}{|s_{12}-m_{X}^2+im_X\Gamma_X|^2}\left[ (s_{12}-m_X^2)\Re (a_S^\ast\tilde{a}_P)+m_X\Gamma_X \Im (a_S^\ast\tilde{a}_P)\right],
\end{equation}
where the first term implies a sign flip at $s = m_{X}^2$.
Roughly speaking, when the phases of $a_S$ and $\tilde{a}_P$ are about the same, the first term would dominate, and there will be large cancellation when $s_{12}$ is integrated from $(m-\sigma_X)^2$ to $(s_{12}+\sigma_X)^2$, resulting in a much smaller FBI-$C\!P$A or regional $C\!P$A for the region $(m_\rho-\sigma_\rho)^2<s_{12}<(m_\rho+\sigma_\rho)^2$.
In fact, the this kind of behaviour has already been observed by LHCb in the decay $B^\pm\to\pi^+\pi^-\pi^\pm$, in which it is shown that the regional $C\!P$ asymmetries for $s_{12}$ below and above the $\rho^{0}(770)$ mass squared, $(m_\rho-\sigma_\rho)^2<s_{12}<m_\rho^2$ and $m_\rho^2<s_{12}<(m_\rho+\sigma_\rho)^2$, tend to take opposite signs \cite{Aaij:2019hzr,Aaij:2019jaq}.
The aforementioned cancellation can be circumvented by choosing different interval of $s_{12}$.
For example, for the case of $B^\pm\to\pi^+\pi^-\pi^\pm$, one can measure the FBI-$C\!P$As defined on $(m_\rho-\sigma_\rho)^2<s_{12}<m_\rho^2$ and $m_\rho^2<s_{12}<(m_\rho+\sigma_\rho)^2$, respectively, instead of that of the combined interval $(m_\rho-\sigma_\rho)^2<s_{12}<(m_\rho+\sigma_\rho)^2$.

Besides the aforementioned motivation, another important advantage of FBI-$C\!P$A is that it can almost effectively double the statistics in the experiments.
Consequently, as a complement to the reginal $C\!P$ asymmetries, FBI-$C\!P$A can be used in searching for $C\!P$ violations in some three-body decays of beauty or charmed meson, in which the $C\!P$ violations are expected to be small so that higher statistics is essential.
To see this, one first needs to notice that FBI-$C\!P$A can be approximated to an experimentally useful expression, which is
\begin{equation}\label{eq:smallCParppox}
  A^{FB}_{C\!P}\approx \frac{\left[\Gamma_{H}(c_{\theta^\ast}>0)+\Gamma_{{\bar H}}(c_{\theta^\ast}<0)\right]-\left[\Gamma_{H}(c_{\theta^\ast}<0)+\Gamma_{{\bar H}}(c_{\theta^\ast}>0)\right]}{\left[\Gamma_{H}(c_{\theta^\ast}>0)+\Gamma_{{\bar H}}(c_{\theta^\ast}<0)\right]+\left[\Gamma_{H}(c_{\theta^\ast}<0)+\Gamma_{{\bar H}}(c_{\theta^\ast}>0)\right]},
  \end{equation}
if the $C\!P$ violation is small.
By comparing to the conventionally defined regional $C\!P$ asymmetries $A^{{\rm reg}}_{C\!P}(c_{\theta^\ast_{1}}\gtrless 0)$ in Eq. (\ref{regionACP1part}), it can be clearly seen that the statistics has indeed almost doubled.

It would be useful to further compare FBI-$C\!P$A with the regional $C\!P$ asymmetry $A^{{\rm reg}}_{C\!P}(c_{\theta^\ast}\!\!<\! 0~\&~c_{\theta^\ast}\!\!>\!0)$, which is defined as
\begin{eqnarray}
&&A^{{\rm reg}}_{C\!P}(c_{\theta^\ast_{1}}\!\!<\! 0~\&~c_{\theta^\ast_{1}}\!\!>\!0)\nonumber\\
&=& \frac{\left[\Gamma_{H}(c_{\theta^\ast_{1}}>0)+\Gamma_{H}(c_{\theta^\ast_{1}}<0)\right]-\left[\Gamma_{{\bar H}}(c_{\theta^\ast_{1}}<0)+\Gamma_{{\bar H}}(c_{\theta^\ast_{1}}>0)\right]}{\left[\Gamma_{H}(c_{\theta^\ast_{1}}>0)+\Gamma_{H}(c_{\theta^\ast_{1}}<0)\right]+\left[\Gamma_{{\bar H}}(c_{\theta^\ast_{1}}<0)+\Gamma_{{\bar H}}(c_{\theta^\ast_{1}}>0)\right]}.
\end{eqnarray}
Just as the cases of $B^\pm\to \pi^+\pi^-\pi^\pm$ and $B^\pm\to\pi^+\pi^- K^\pm$,  although the regional $C\!P$ asymmetries $A^{{\rm reg}}_{C\!P}(c_{\theta^{\ast}_{1}}>0)$ and $A^{{\rm reg}}_{C\!P}(c_{\theta^{\ast}_{1}}< 0)$  around the vicinity of $\rho^0$ can be large, they tend to take opposite signs because of the interference of the $S$- and $P$-waves, hence there is cancellation when summing up the event yields to obtain the regional $CP$ asymmetry $A^{{\rm reg}}_{C\!P}(c_{\theta^\ast_{1}}\!\!<\! 0~\&~c_{\theta^\ast_{1}}\!\!>\!0)$.
In fact, the $C\!P$ asymmetry originated from the interference of the $S$- and $P$-waves is totally cancelled, which can be seen from the expression:
\begin{equation}
A^{{\rm reg}}_{C\!P}(c_{\theta^\ast_{1}}\!\!<\! 0~\&~c_{\theta^\ast_{1}}\!\!>\!0)=\frac{A_{C\!P}^P+3\frac{\left\langle\left| {a}_S\right|^2\right\rangle+\left\langle\left| {\bar a}_S\right|^2\right\rangle}{\left\langle\left| {a}_P\right|^2\right\rangle+\left\langle\left| {\bar a}_P\right|^2\right\rangle}A_{CP}^S}{1+3\frac{\left\langle\left| {a}_S\right|^2\right\rangle+\left\langle\left| {\bar a}_S\right|^2\right\rangle}{\left\langle\left| {a}_P\right|^2\right\rangle+\left\langle\left| {\bar a}_P\right|^2\right\rangle}}.
\end{equation}
Consequently, FBI-$C\!P$A may take larger values than $A^{{\rm reg}}_{C\!P}(c_{\theta^\ast_{1}}\!\!<\! 0~\&~c_{\theta^\ast_{1}}\!\!>\!0)$, making the former easier to observe.

From the above analysis, one can see that FBI-$C\!P$A serves as a complementary observable for $C\!P$ asymmetries around the vicinity of $X$, along with $A^{{\rm reg}}_{C\!P}(c_{\theta^{\ast}_{1}}>0)$, $A^{{\rm reg}}_{C\!P}(c_{\theta^{\ast}_{1}}<0)$, and $A^{{\rm reg}}_{C\!P}(c_{\theta^\ast_{1}}\!\!<\! 0~\&~c_{\theta^\ast_{1}}\!\!>\!0)$.
Moreover, for some multi-body decays of $B$ and $D$ mesons, it has a good chance that $C\!P$ violation could be first confirmed through the measurement of FBI-$C\!P$A.

\section{\label{sec:appl}Application potential to multi-body decays of charm and beauty mesons}

There are a lot of channels which are suitable to perform the measurements of FBI-$C\!P$A.
In the $B$ meson sector, for channels such as $B^\pm\to \pi^+\pi^-K^\pm $ and $B^\pm\to \pi^+\pi^-\pi^\pm $ \cite{Aaij:2014iva}, there are very clear interference effect between $S$- and $P$-wave when the invariant mass of $\pi^+\pi^-$ lies around the vicinity of the vector resonance $\rho^0(770)$.
The regional $C\!P$ asymmetries has already been measured by LHCb.
We suggest to perform measurements of FBI-$C\!P$A around $\rho^0(770)$ in these channels.
For channels such as $B^\pm \to K^+K^- K^\pm$ or $B^\pm\to K^+K^-\pi^\pm$ \cite{Aaij:2014iva}, FBI-CPV around the $P$-wave resonances such as $\phi(1020)$ are also worth measuring.

Similarly, measurements of FBI-$C\!P$A could potentially find evidence of $C\!P$ violations in $D^\pm\to K^+K^-\pi^\pm$ \cite{Aaij:2011cw} and $D^\pm_{(s)}\to\pi^+\pi^-\pi^\pm$ \cite{Aaij:2013jxa}.
For $D^\pm\to K^+K^-\pi^\pm$, the resonances $K^\ast(892)$ and $\phi(1020)$ are clearly visible in the Dalitz plot. The forward-backward asymmetries for these two $P$-wave resonances are also visible.
It would be interesting to check weather the $CP$ violation shows up in FBI-$C\!P$A around these resonances.
For $D^\pm\to\pi^+\pi^-\pi^\pm$, the vector resonance $\rho^0(770)$ and its forward-backward asymmetry is also visible.

For an illustration, we consider the decay process $D^\pm\to K^+K^-\pi^\pm$.
From FIG. 2 of Ref. \cite{Aaij:2011cw} one can see that the forward-backward asymmetry around the resonance $\overline{K}^\ast(892)^0$ is quite clear, indicating an interference effect.
This interference is probably caused by the S-wave resonance $\overline{K}_0^\ast(700)$.
If this is the case, the decay amplitudes in the phase space region around the vicinity of the resonance $\overline{K}^\ast(892)^0$ can then be expressed as
\begin{equation}
  \mathcal{M}=\frac{1}{s_{\overline{K}^\ast}}\mathcal{M}_{D^+\to K^+ \overline{K}^\ast}\mathcal{M}_{\overline{K}^\ast\to K^-\pi^+}+\frac{1}{s_{\overline{K}_0^\ast}}\mathcal{M}_{D^+\to K^+ \overline{K}_0^\ast}\mathcal{M}_{\overline{K}_0^\ast\to K^-\pi^+}
\end{equation}
where $s_{X}=s-m_X^2+i m_X\Gamma_X$ ($X=\overline{K}^\ast$ or $\overline{K}^\ast_0$), and the decay amplitudes can be parameterized as
\begin{equation}
 \mathcal{M}_{D^+\to K^+ \overline{K}^\ast} \mathcal{M}_{\overline{K}^\ast\to K^-\pi^+}=c_{\theta^\ast}\left(\lambda_s\eta_s+\lambda_d\eta_d+\lambda_b\eta_b\right),
\end{equation}
\begin{equation}
 \mathcal{M}_{D^+\to K^+\overline{K}^\ast_0} \mathcal{M}_{\overline{K}_0^\ast\to K^-\pi^+}=\lambda_s\xi_s+\lambda_d\xi_d+\lambda_b\xi_b,
\end{equation}
where $\lambda_q=V_{uq}V_{cq}^\ast$.
Since $\lambda_s\gg \lambda_b$, we can rewrite the above two amplitudes into the form
\begin{equation}
 \mathcal{M}_{D^+\to K^+ \overline{K}^\ast} \mathcal{M}_{\overline{K}^\ast\to K^-\pi^+}=c_{\theta^\ast}\left(\lambda_s\tilde{\eta}_s+\lambda_b\tilde{\eta}_b\right),
\end{equation}
\begin{equation}
 \mathcal{M}_{D^+\to K^+\overline{K}^\ast_0} \mathcal{M}_{\overline{K}_0^\ast\to K^-\pi^+}=\lambda_s\tilde{\xi}_s+\lambda_b\tilde{\xi}_b,
\end{equation}
where $\tilde{\eta}_{s/b}=\eta_{s/b}-\eta_{d}$, $\tilde{\xi}_{s/b}=\xi_{s/b}-\xi_{d}$, and $\phi$ is the phase difference between $\lambda_s$ and $\lambda_b$.
The FBI-$C\!P$A is then approximated to be
\begin{equation}
  A_{C\!P}^{FB}\approx\frac{6\Im\left[\left(\frac{\tilde{\xi}_b}{\tilde{\eta}_s}-\frac{\tilde{\eta}_b\tilde{\xi}_s}{\tilde{\eta}_s^2} \right)^\ast\left\langle\frac{1}{s_{K^\ast}s_{K^\ast_0}^\ast}\right\rangle\right] }{\left\langle\left|\frac{1}{s_{K^\ast}}\right|^2\right\rangle}\left|\frac{\lambda_b}{\lambda_s}\right|\sin\phi,
\end{equation}
The relative strong phase between the amplitudes corresponding to these two resonances can be large because of the non-perturbative effect.
As a consequence, FBI-$C\!P$A will be roughly of the order $ A_{C\!P}^{FB}=|\frac{\lambda_b}{\lambda_s}|\sin\phi\sim0.1\%$, which is just about the same order with the regional $C\!P$As.
In order to distinct from zero for such a small value, the statistics should be large enough.
In this sense, the measurement of FBI-$C\!P$A is better than that of the regional $C\!P$As, as the former can make use of the data more efficiently.
Although FBI-$C\!P$A is defined through forward-backward asymmetry, one does not need to obtain FBI-$C\!P$A by means of the measurement of forward-backward asymmetry at all for this situation.
Since the CP asymmetry is quite small, one just needs to measure FBI-$C\!P$A in $D^\pm\to K^+K^-\pi^\pm$ around $\overline{K}^\ast(892)^0$ according to Eq. (\ref{eq:smallCParppox}), from which
one can see that the statistics are indeed almost doubled comparing to the regional $C\!P$ asymmetry in Eq. (\ref{regionACP1part}).

In fact, besides the above suggested decay channels, both the measurements of the forward-backward asymmetry and FBI-$C\!P$A are meaningful in other multi-body decay channels of charm and beauty meson, provided that a $P$-wave resonances is presented in the Dalitz plot.

\section{\label{sec:concl}Conclusion}

To sum up, we introduce an observable for $C\!P$ violations in multi-body decays of heavy meson, the forward-backward asymmetry induced $C\!P$ asymmetry, FBI-$C\!P$A.
We suggest to perform the measurements of FBI-$C\!P$A in some decay channels of charm and beauty mesons.

\section*{Acknowledgments}
 This work was supported by National Natural Science Foundation of China under Contracts No. 11705081.

\bibliographystyle{elsarticle-num}
\bibliography{zzh}





\end{document}